\documentclass[oneside,reqno,a4paper,12pt]{amsart}

\usepackage{latexsym}
\usepackage{euscript}
\usepackage{epsfig}

\newcommand{\be}{\begin{equation}}
\newcommand{\ee}{\end{equation}}

\def\eg{{\it e.g.}}
\def\ie{{\it i.e.}}
\def\LEP2{{LEPII}}

\begin{document}
\begin{flushright}
BA-99-44
\end{flushright}
\vskip 1cm
\title[]{Supersymmetric CP violation $\varepsilon'/\varepsilon$
due to asymmetric\\ \vskip 0.2cm  $A$-matrix}

\maketitle

\begin{center}
\textsc{Shaaban Khalil$^{1,2}$ and Tatsuo Kobayashi$^{3}$}
\\ \vspace*{1cm} \small{\textit{$^1$Bartol Research Institute,
University of Delaware Newark, DE 19716}} \\ \vspace*{3mm}
\small{\textit{$^2$Ain Shams University, Faculty of Science, Cairo
11566, Egypt}} \\ \vskip 0.3cm \small{\textit{$^3$Department of
Physics, High Energy Physics Division, University of Helsinki\\
and \\ Helsinki Institute of Physics, P.O. Box 9
(Siltavuorenpenger 20 C)\\ FIN-00014 Helsinki, Finland }}\\
\end{center}
\vspace*{0.8cm}
\begin{center}
ABSTRACT
\end{center}
\vspace*{5mm}
\renewcommand{\baselinestretch}{1.3} \large\normalsize
\begin{quotation}
We study contributions of supersymmetric CP phases to the CP
violation $\varepsilon'/\varepsilon$ in models with asymmetric
$A$-matrices. We consider asymmetric $A$-matrices, which are
obtained from string-inspired supergravity. We show that a
certain type of asymmetry of $A$-matrices enhances supersymmetric
contributions to the CP violation $\varepsilon'/\varepsilon$ and
the supersymmetric contribution to $\varepsilon'/\varepsilon$ can
be of order of the KTeV result, $\varepsilon'/\varepsilon \sim
10^{-3}$.
\end{quotation} \vspace*{9mm}
{\bf 1.} CP violation is sensitive to physics beyond the standard
model (SM), that is, CP violation , as well as flavor changing
neutral current (FCNC) processes, constrains significantly physics
beyond the SM, \eg, supersymmetric models. On the other hand, if
any deviation of CP violation from the SM is observed, that would
be a strong hint to physics beyond the SM. Many works have been
done on CP violation in supersymmetric models
\cite{cpsusy,epsk,masiero,cprev}. \vskip 0.3cm Recently the KTeV
collaboration at Fermilab has reported a measurement of the
direct CP-violation in $K \to \pi \pi$ decays~\cite{ktev} $$
\mathrm{Re} (\varepsilon'/\varepsilon) = (28 \pm 4) \times
10^{-4}.$$ This measurement has confirmed the previous result of
the NA31 experiment at CERN \cite{na31}. Hence, it excludes the
superweak models. The SM predicts non-zero value for
$\varepsilon'/\varepsilon$. However, its prediction suffers from
a large ambiguity due to the theoretical uncertainties in the
hadronic quantities. Recent discussions on
$(\varepsilon'/\varepsilon)$ can be found in
Refs.\cite{keum}-\cite{benatti}. \vskip 0.3cm In supersymmetric
models, there are several possible sources for CP violation
beyond the Cabibbo-Kobayashi-Maskawa (CKM) phase $\delta_{CKM}$
in the SM. Two types of physical phases only remain in the minimal
supersymmetric standard model (MSSM) after all appropriate field
redefinition, namely, the phases of $A$-parameters and the phase
of the $\mu$-term ($\phi_{\mu}$). A generic type of $A$-matrices
include many degree of freedom for the real part and imaginary
part. However, in most of cases the universality of $A$-matrices
has been assumed. That is good to simplify calculations, but that
removes interesting degree of freedom, in particular for CP
violation. \vskip 0.3cm Several implications of non-universal
$A$-matrices have been discussed in Ref.~\cite{abel,khalil}. The
non-universality among the soft supersymmetry (SUSY) breaking
terms plays an important role on all CP violating processes. In
particular, it has been shown in Ref.~\cite{khalil} that
non-degenerate $A$-parameters can generate the experimentally
observed CP violation $\varepsilon$ even with the vanishing CKM
CP phase $\delta_{CKM}=0$, that is, the fully supersymmetric CP
violation in the Kaon system is possible. In Ref.~\cite{khalil},
an example of symmetric $A$-matrices for the first $2 \times 2$
block has been considered with the exact degeneracy of squark
masses between the first and second families. Because
string-derived soft terms \cite{BIM,KSYY} require symmetric
$A$-matrices  for exactly degenerate soft masses. This model can
lead to the parameter region  where we have the SUSY contribution
of $O(10^{-3})$ to $\varepsilon$. However, this type of models
provide with a very small value for $\varepsilon'/\varepsilon$.
This result is due to an accidental cancellation between the
different contributions because of the symmetric form of the
$A$-matrices which have been used. \vskip 0.3cm In this letter,
we study the SUSY contributions to the CP violation
$\varepsilon'/\varepsilon$ in models with asymmetric $A$-matrices.
We consider two possibilities for asymmetric $A$-matrices keeping
the degeneracy of squark masses. One model has almost degenerate
squark masses, which are realized by dilaton-dominate SUSY
breaking. In the other model, we require a delicate cancellation
between string-derived soft masses and the $D$-term contributions
to soft masses. The latter can lead to a large asymmetry for the
$A$-matrix. Using these models, we calculate
$\varepsilon'/\varepsilon$ explicitly. Then we show that in the
case with asymmetric $A$-terms the SUSY contribution to
$\varepsilon'/\varepsilon$ can be of order of the KTeV result,
$\varepsilon'/\varepsilon \sim 10^{-3}$. In the whole analysis we
take $\delta_{CKM}=0$ in order to show the pure SUSY
contributions. \vskip 0.3cm

{\bf 2.} Here we consider the possibilities that  one can obtain
non-degenerate $A$-matrices keeping degeneracy of the squark
masses. First we give a brief review on the soft SUSY breaking
terms in string models,
\begin{eqnarray}
-\mathcal{L}_{\mathrm{SB}} &=& \frac{1}{6} \,h_{ijk}\,\phi_i
\phi_j \phi_k + \frac{1}{2} \,(\mu B)^{ij}\,\phi_i \phi_j +
\frac{1}{2} \,(m^2)^{j}_{i}\,\phi^{*\,i} \phi_j+ \frac{1}{2}
\,M_a\,\lambda \lambda+\mbox{H.c.}~
\end{eqnarray}
where the $\phi_i$ are the scalar parts of the chiral superfields
$\Phi_i$ and $\lambda$ are the gauginos. We use the notation of
trilinear coupling terms, the so-called $A$-terms, as
$h_{ijk}=(YA)_{ijk}$, where $Y_{ijk}$ is the corresponding Yukawa
coupling. \vskip 0.3cm We start with the (weakly coupled)
string-inspired supergravity theory. Its  K\"ahler potential is
\begin{equation}
K= -\ln (S+S^*)+ 3 \ln (T+T^*) +\sum_i (T+T^*)^{n_i}|\Phi^i|^2,
\end{equation}
where $S$ and $T$ are the dilaton field and the moduli field.
Now we assume a nonperturbative superpotential of $S$ and $T$,
$W_{np}(S,T)$, is induced and $F$-terms of $S$ and $T$ contribute
to SUSY breaking. In addition, we assume the vanishing vacuum
energy. Then we parameterize $F$-terms \cite{BIM}
\begin{eqnarray}
F^S=\sqrt 3 m_{3/2}(S+\bar S)\sin \theta e^{-i\alpha_S},\  F^T=
m_{3/2}(T+\bar T)\cos \theta e^{-i\alpha_T}.
\end{eqnarray}
Within this framework, the soft scalar mass and the $A$-parameter
are obtained
\begin{eqnarray}
m^2_i &=& m^2_{3/2}(1 + n_i \cos^2\theta),
\label{scalar}\\
A_{ijk} &=& - \sqrt{3} m_{3/2} \sin\theta e^{-i \alpha_s}
- m_{3/2} \cos\theta
(3 + n_i + n_j + n_k) e^{-i \alpha_T},
\label{trilinear}
\end{eqnarray}
where $n_i$, $n_j$ and $n_k$ are modular weights of fields in the
corresponding Yukawa coupling $Y_{ijk}$.
Here we have assumed the corresponding Yukawa coupling $Y_{ijk}$
is $T$-independent.
In addition, the gaugino masses are obtained
\begin{eqnarray}
M_a &=& \sqrt{3} m_{3/2} \sin\theta e^{- i \alpha_{S}}.
\label{gaugino}
\end{eqnarray}

It is obvious that if we require exact degeneracy of squark masses
between the first and second families, i.e. $n_{D1}=n_{D2}$ and
similar relations for the other squarks, the first $2 \times 2$
blocks of the $A$-matrices for the up and down sector, $A^u_{ij}$
and $A^d_{ij}$, are degenerate. That is what have been used in
Ref.~\cite{khalil}. In such a case we obtain a suppressed
contribution to the CP violation $\varepsilon'/\varepsilon$.
\vskip 0.3cm In the first model we use, we assign different
modular weights for the first and second families in order to have
asymmetric $A$ matrices. In the dilaton-dominant case with $\tan
\theta >> 1$, we have almost degeneracy of the squark masses. In
addition, the renormalization group effects due to the gaugino
masses dilute the nondegeneracy. In Ref.~\cite{BIM} it has been
shown that the goldstino angle $\theta$ is constrained $\cos^2
\theta <1/3$ for $n_i-n_j=1$ from the FCNC\footnote{Furthermore,
in Ref.~\cite{louis} it has been shown that a certain type of
non-universal $A$-terms reduce the difference at low energy and
even could tune it to vanish.}. For example, as our first model
with the asymmetric $A$-matrix, we take the following assignment
of the modular weights
\begin{eqnarray}
& n_{Q_1}=-1, \hspace{1cm} n_{Q_2}=-2, \hspace{1cm} n_{Q_3}=-3,\nonumber \\
& n_{Di}=n_{Ui}=n_{H_1}=-1, \hspace{1cm} n_{H_2} = -3,
\end{eqnarray}
where $i=1,2,3$. On the top of that, we restrict our analysis to
the region $\cos^2 \theta <1/3$. \vskip 0.3cm Under this
assumption, we have the $A$-parameter matrix for the down sector,
\begin{eqnarray}
A^{d}_{ij} = \left (
\begin{array}{ccc}
a_{d} & a_{d} & a_{d} \\
b_{d} & b_{d} & b_{d} \\
c_{d} & c_{d} & c_{d}
\end{array}
\right),
\label{Atex}
\end{eqnarray}
where
\begin{eqnarray}
a_d&=& -\sqrt{3} m_{3/2} \sin\theta, \nonumber\\
b_d&=& m_{3/2}(-\sqrt{3} \sin\theta +  e^{-i \alpha'} \cos\theta),\\
c_d&=& m_{3/2}(-\sqrt{3} \sin\theta + 2 e^{-i \alpha'} \cos\theta)
\nonumber.
\end{eqnarray}
Here we have rotated the gaugino mass terms into real and rotated
the $A$-terms at the same time and $\alpha'$ denotes $\alpha'
\equiv \alpha_T-\alpha_S$. In this case, we have the asymmetry
between $A^d_{12}$ and $A^d_{21}$, but it is limited because of
$\cos^2 \theta <1/3$. Such asymmetry can be enlarged in the
$h_{ijk}$-matrix if we take an asymmetric Yukawa matrix. Thus, we
discuss the two cases: one case has a typical symmetric Yukawa
matrix and the other case has an example of asymmetric Yukawa
matrices. As a typical type of the symmetric and realistic Yukawa
matrices, we use the type which are shown explicitly in
Ref.~\cite{khalil}. The Yukawa matrices among this type lead to
similar results for the CP violation each other. As an example of
asymmetric Yukawa matrices, we take the following form,
\begin{eqnarray}
Y^{u}_{ij} = y^u\left (
\begin{array}{ccc}
\lambda^8 &\lambda^5  & \lambda^3 \\
\lambda^7 & \lambda^4 &\lambda^2  \\
\lambda^5 & \lambda^2 & 1
\end{array}
\right), \quad
Y^{d}_{ij} = y^d \left (
\begin{array}{ccc}
\lambda^5 &\lambda^3  & \lambda^3 \\
\lambda^4 & \lambda^2 &\lambda^2  \\
\lambda^2 & 1 & 1
\end{array}
\right),
\label{asym-Y}
\end{eqnarray}
where $\lambda \sim 0.22$. These correspond to the Yukawa
matrices with one $O(\lambda)$ deviation in Ref.~\cite{dudas}.
\vskip 0.35cm

{\bf 3.} We explain the second model which we use. We assume
an extra $U(1)$ gauge symmetry \footnote{This $U(1)$ symmetry may
be anomalous \cite{au1} or anomaly-free.} and it is broken by the
vacuum expectation value (VEV) of the Higgs field $\chi$. This
breaking induces another type of contribution to soft scalar
masses, i.e. the $D$-term contribution, which is proportional to a
charge of the broken symmetry. In this case, the soft scalar mass
is obtained
\begin{eqnarray}
m^2_i &=& m^2_{3/2}(1 + n_i \cos^2\theta)+q_im_D^2,
\label{scalar2}
\end{eqnarray}
where $q_i$ is the charge of the broken $U(1)$ of the matter field,
and $m_D^2$ is the universal part of the $D$-term contributions.
Thus, the soft scalar masses are, in general, non-degenerate
for $n_i$ and $q_i$.
However, the soft scalar masses $m_i^2$ and $m_j^2$ are degenerate
if the following two conditions are satisfied,

(a) $n_i-n_j = C(q_i - q_j)$,\\
where $C$ is universal for $i$ and $j$,

(b) $m^2_{3/2} \cos^2\theta + m_D^2/C =0$.\\
In this case, we obtain the degenerate soft scalar masses for
different $n_i$ and $n_j$,
that is, we can obtain non-degenerate $A$-matrices keeping degenerate
soft scalar masses.
Thus, this is a very interesting fine-tuned case in the whole
parameter space.

Before calculations of
$\varepsilon'/\varepsilon$ in this model, we give comments on the
conditions (a) and (b). We denote here the modular weight and the
$U(1)$ charge of $\chi$ by $n_\chi$ and $q_\chi$ and we take the
normalization such that $q_\phi = -1$. The VEV of $\chi$ induces
the following terms in the superpotential,
\begin{eqnarray}
W_{Yukawa}
&=& Y^u_{ij}\theta(q_{H2}+q_{Qi}+q_{Uj})(<\chi> / M)^{(q_{H2}+q_{Qi}+q_{Uj})}
Q^iU^jH_2
\nonumber \\
&+& Y^d_{ij}\theta(q_{H1}+q_{Qi}+q_{Dj})(<\chi> / M)^{(q_{H1}+q_{Qi}+q_{Dj})}
Q^iD^jH_1.
\end{eqnarray}
The superpoetential includes a similar term for the lepton sector.
Here the couplings $Y^u_{ij}$ and $Y^d_{ij}$ are naturally of
$O(1)$. The suppression factors $(<\chi> /
M)^{(q_{H2}+q_{Qi}+q{Uj})}$ and $(<\chi> /
M)^{(q_{H1}+q_{Qi}+q{Dj})}$ can lead to realistic hierarchies of
the Yukawa matrices \cite{FN,FN2,dudas}. \vskip 0.3cm Now we
consider the $T$-duality transformation,
\begin{equation}
T \rightarrow {aT -ib \over icT + d},
\end{equation}
where $ad-bd=1$ and $a,\cdots,d$ are integers.
We assume that the chiral field transforms
\begin{equation}
\Phi^i \rightarrow (icT+d)^{n_i} \Phi^i.
\end{equation}
Then we require $G \equiv K+\ln |W|^2$ is duality-invariant.
That implies that the superpotential should have the total modular weight
$\sum n_i =-3$, that is, \cite{Dterm2}\footnote{See also
for $D$-term contributions derived superstring theory e.g.
\cite{Dterm3}.}
\begin{eqnarray}
(q_{H2}+q_{Qi}+q_{Uj})n_\chi+n_{H2}+n_{Qi}+n_{Uj} &=& -3, \nonumber \\
(q_{H1}+q_{Qi}+q_{Dj})n_\chi+n_{H1}+n_{Qi}+n_{Dj} &=& -3,
\end{eqnarray}
for non-vanishing couplings.
Thus, it is obvious that for $Q^i$ fields the condition (a) is satisfied,
i.e.
\begin{equation}
n_{Qi}-n_{Qj}=- n_\chi (q_{Qi} - q_{Qj}),
\end{equation}
if the $(i,k)$ and $(j,k)$ entries do not vanish for one of $k$ in
either the up-sector or down-sector.
Similarly we obtain
\begin{equation}
n_{Ui}-n_{Uj}=- n_\chi (q_{Ui} - q_{Uj}), \quad
n_{Di}-n_{Dj}=- n_\chi (q_{Di} - q_{Dj}),
\end{equation}
if $(k,i)$ and $(k,j)$ entries do not vanish for one of $k$ in
the up-sector and down-sector, respectively. Thus, a certain type
of symmetries can realize the condition (a). \vskip 0.3cm Let us
give a comment on the condition (b), too. Now our free parameters
are $m_{3/2}$, $\theta$ and $m_D^2$ and these are determined by
VEVs. Note that if the condition (a) is satisfied, we have the
very special direction in our parameter space, i.e.
\begin{equation}
m_{3/2}\cos^2\theta =n_{\chi}m_D^2,
\label{dir}
\end{equation}
which corresponds to the condition (b). Along this direction, we
have the degenerate soft scalar masses, $m_{Q(U,D)i}=m_{Q(U,D)j}$
for $i,j$. For simplicity, we assume that all squark masses are
universal for the direction (\ref{dir}), i.e.
$m_{Q(U,D)i}^2=m_0^2$. Now we consider a vicinity around
(\ref{dir}) and there are two types of directions to change
parameters. One is to violate the degeneracy and the other is to
keep the degeneracy. Here let us consider only the direction to
violate the degeneracy and treat the degree of freedom of the
corresponding VEV as a dynamical parameter. Thus, we can write
\begin{equation}
m_i^2 = m_0^2+q_i\delta m_0^2,
\end{equation}
around (\ref{dir}), where $\delta m_0^2$ corresponds to the
dynamical direction to violate the degeneracy. It is obvious that
such direction is proportional to $q_i$ for $m_i$, because of
eq.(\ref{scalar2}) and the linear relation between $q_i$ and
$n_i$. \vskip 0.cm Now let us consider the one-loop effective
potential,
\begin{equation}
\Delta V_1 = {1 \over 64 \pi^2}\mathrm{Str}\mathcal{M}^4 (\ln
\mathcal{M}^2/Q^2 -{3/2}).
\end{equation}
Around (\ref{dir}), we can expand
\begin{eqnarray}
 \Delta V_1 &=& [\Delta V_1]_{m_i^2=m_0^2}+
\mathrm{Tr~} q_i [d \Delta V_1 /d \delta
m_0^2]_{m_i^2=m_0^2}\delta m_0^2 \nonumber \\ &+& \mathrm{Tr~}
(q_i)^2 [d^2 \Delta V_1 /d (\delta m_0^2)^2]_{m_i^2=m_0^2} (\delta
m_0^2)^2 +\cdots.
\end{eqnarray}
Note that the gaugino masses have no direction corresponding to
$\delta m_0^2$. If $\mathrm{Tr~} q_i=0$ and $d^2 \Delta V_1 /d
(\delta m_0^2)^2 > 0$, the fine-tuning direction (\ref{dir}) is a
locally minimum. \vskip 0.3cm As explained above, we obtain the
degenerate soft scalar mass with non-degenerate $A$-parameters
under the conditions (a) and (b). Note that the assignment of the
modular weights are related with the assignment of the $U(1)$
charges. As a simple example, we use the following assignment,
\begin{equation}
n_{Q_1}=-4, \hspace{1cm} n_{Q_2}=-3, \hspace{1cm} n_{Q_3}=-1,
\end{equation}
\begin{equation}
n_{U_1}=-6, \hspace{1cm} n_{U_2}=-3, \hspace{1cm} n_{U_3}=-1,
\end{equation}
\begin{equation}
n_{D_1}=-3, \hspace{1cm} n_{D_2}=-1, \hspace{1cm} n_{D_3}=-1,
\end{equation}
\begin{equation}
n_{H_1}=-1, \hspace{1cm} n_{H_2}=-1.
\end{equation}
This assignment of the modular weights and the corresponding
$U(1)$ charge assignment lead to the Yukawa matrices
(\ref{asym-Y}) \cite{dudas}. In realistic string models, we have
constraints on the modular weights for the MSSM matter fields
\cite{lust}, and it might be difficult to obtain e.g. the quark
field with $n_{U_1}=-6$. However, we use this assignment as a toy
supergravity model. If we make our model more complicated e.g. by
use of complicated extra symmetries or we concentrate only the
first two families, it might be possible to assign more natural
values of modular weights. \vskip 0.3cm Namely, our initial
conditions in the second model are as follows. We assume
$m_i^2=m_0^2$, and furthermore we assume $m_0^2=m_{3/2}^2$. We
use the Yukawa matrices (\ref{asym-Y}) and the $A$-matrices
(\ref{trilinear}) with the above assignment of the modular
weights. \vskip 0.35cm

{\bf 4.} The convenient basis to discuss flavor changing effects
in the SUSY loop with gluino exchange is the so-called super-CKM
basis~\cite{hall}. In this basis the relevant quark mass matrix is
diagonalized, and the squarks are rotated in the same way. We
define $(\delta_{ij})_{LR}$ by normalizing the off diagonal
components by average squark mass squared $m_{\tilde{q}}^2$. The
relevant contribution, in our models, to CP violation comes from
the terms proportional to $(\delta_{12})_{LR}$ and
$(\delta_{12})_{RL}$. The  mass insertion $(\delta_{12})^d_{LR}$,
for instance, is given by
\begin{equation}
(\delta_{12})^d_{LR} = { 1 \over m_{\tilde{q}}^2} U_{1i}
(h^{d})_{ij} U^T_{j2},
\label{delta}
\end{equation}
where $U$ is the matrix diagonalizing the down quark mass matrix.
It was shown in Ref.~\cite{khalil} that to obtain a large SUSY
contribution to $\varepsilon$ it is necessary to enhance the
values of $\mathrm{Im} (\delta_{12})_{LR}$ and $\mathrm{Im}
(\delta_{12})_{RL}$. The non-degeneracy of $A$-terms is an
interesting example for enhancing these quantities. From
eq.(\ref{delta}) we notice that the phases of the all entries of
the $A$-matrix contribute to $(\delta_{12})_{LR}$.
It is remarkable that we have $\mathrm{Im}
(\delta^d_{12})_{LR} \neq \mathrm{Im} (\delta^d_{12})_{RL}$ in
the models with asymmetric $h$-matrices unlike the case of
symmetric $h$-terms. For a wide region of the parameter space we
find that $\mathrm{Im}(\delta^d_{12})_{LR}$ is of order
$10^{-4}-10^{-5}$. That is similar to the case with the symmetric
$h$-terms discussed in Ref.~\cite{khalil}. For these values the
CP violation $\varepsilon$, as explained in Ref.~\cite{khalil},
is of order of the experimental limit $2.2 \times 10^{-3}$. Note
that this is a genuine SUSY contribution since we are assuming
that $\delta_{CKM}=0$. Also in this analysis we have assumed the
vanishing of the phase of $\phi_{\mu}$, which is usually
constrained by the experimental limit of the electric dipole
moment of the neutron~\cite{ibrahim, barr}. \vskip 0.3cm

Now we estimate the $\Delta S=1$ CP violating parameter
$\varepsilon'/ \varepsilon$ in our two models.
$(\delta^d_{12})_{LR}$ and $(\delta^d_{12})_{RL}$ give the
important contributions to the CP violation processes in kaon
physics. The relevant part of  the effective hamiltonian
$\mathcal{H}_{\mathrm{eff}}$ for $\Delta S=1$  CP violation is
\begin{equation}
\mathcal{H}_{\mathrm{eff}} = C_8 O_8 + \tilde{C}_8 \tilde{O}_8,
\end{equation}
where the Wilson coefficient $C_8$ and the $\Delta S=1$ effective
Hamiltonian $O_8$ are given in Ref.\cite{masiero}. The dependence
on $(\delta^d_{12})_{LR}$ and $(\delta^d_{12})_{RL}$ appear in
$C_8$ and $\tilde{C}_8$,
\begin{equation}
C_8 = \frac{\alpha_s \pi}{m_{\tilde{q}}^2} \left[
(\delta^d_{12})_{LL} (-\frac{1}{3} M_3(x) -3M_4(x)) +
(\delta^d_{12})_{LR} \frac{m_{\tilde{g}}}{m_s}(-\frac{1}{3} M_1(x)
-3M_2(x)) \right]
\end{equation}
where $x=m_{\tilde{g}}^2/m_{\tilde q}^2$
and $\tilde{C}_8$ can be obtained from $C_8$ by exchange
$L\leftrightarrow R$ while the matrix element of the operator
$\tilde{O}_8$ is obtained from the matrix element of $O_8$
multiplying them by $(-1)$. The functions $M_1$, $M_2$, $M_3$ and
$M_4$ can be found in Ref.~\cite{masiero}.
The expression for $\varepsilon'$ is given by
the following formula,
\begin{equation}
\varepsilon' = e^{\frac{\pi}{4}} \frac{\omega}{\sqrt{2}} \xi
(\Omega -1)
\end{equation}
where $\omega= \mathrm{Re} A_2/\mathrm{Re} A_0$, $\xi =\mathrm{Im}
A_0/ \mathrm{Re} A_0$ and $\Omega = \mathrm{Im} A_2 / (\omega
\mathrm{Im} A_0)$. The amplitudes $A_I$ are defined as $\langle
\pi \pi(I) \vert \mathcal{H}_{eff} \vert K^0 \rangle$ where
$I=0,2$ is the isospin of the final two-pion state. \vskip 0.3cm
Let us discuss the first model with the symmetric and asymmetric
Yukawa matrices. We take $\cos \theta =1/\sqrt{10}$ and
$\alpha'\simeq \pi/2$. In both cases, we have
$\mathrm{Im}(\delta^d_{12})_{LR}$ and
$\mathrm{Im}(\delta^d_{12})_{RL}$ of order $10^{-4} - 10^{-5}$
and they are not degenerate. Thus, it is possible to obtain large
values of $\varepsilon'/\varepsilon$. Indeed for
$\mathrm{Im}(\delta^d_{12})_{LR}\simeq 10^{-5}$ we obtain
$\varepsilon'/\varepsilon$ of order of the experimental results of
NA31 and KTeV, $\varepsilon'/\varepsilon =2.8\times 10^{-3}$. It
is interesting to note that even for
$\mathrm{Im}(\delta^d_{12})_{LR} \sim 10^{-5}$ one can obtain a
sizable contribution to $\varepsilon$. As pointed out in
Ref.~\cite{masiero2},  for $\mathrm{Re}(\delta^d_{12})_{LR}
\simeq 10^{-3}$ the $\varepsilon$-requirement $\vert 2
\mathrm{Re}(\delta^d_{12})^2_{LR}
\mathrm{Im}(\delta^d_{12})^2_{LR} \vert^{\frac{1}{2}} \simeq
2\times 10^{-4} $ is satisfied for the region of the parameter
space which leads to $\varepsilon'/\varepsilon \sim 10^{-3}$.
\vskip 0.3cm Figure 1 shows $\varepsilon'/\varepsilon$ versus
$m_{3/2}$. The solid line and the dotted line correspond to the
symmetric and the asymmetric Yukawa matrices, respectively. In
the both cases, the SUSY phase has a sizable contribution to
$\varepsilon'/\varepsilon$. As expected, the asymmetric Yukawa
matrix provides with larger values of $\varepsilon'/\varepsilon$
than the symmetric case. For example, the asymmetric Yukawa
matrix leads to the value $\varepsilon'/\varepsilon =2.8\times
10^{-3}$ at $m_{3/2} \sim 450$ GeV.
\vskip 0.3cm
\begin{figure}[h]
\psfig{figure=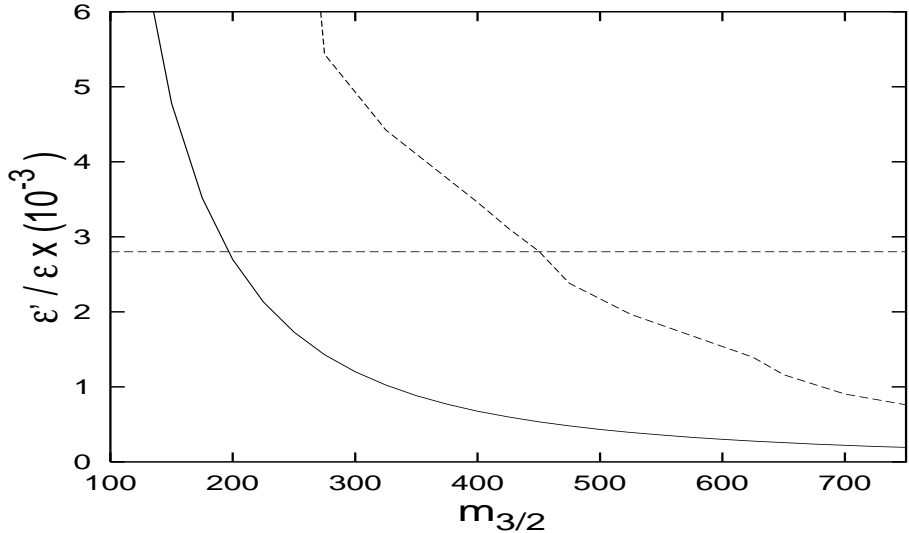,height=7cm,width=12cm} \caption{The
value of $\varepsilon'/\varepsilon$ as a function of the gravitino
mass} \vskip 0.3cm
\end{figure}

Similarly,  we can discuss the second model. Actually, the
asymmetry of the $A$-matrices is large compared with the first
model. Thus, the second model predicts much larger values of
$\varepsilon'/\varepsilon$ for the same value of $\cos \theta$.
For example, we have $\varepsilon'/\varepsilon =O(10^{-2}\sim
10^{-1})$ for $\cos \theta=1/\sqrt{10}$ and $m_{3/2}=O(100)$ GeV.
Such a case, \ie ~such magnitude of asymmetry in the $h$-matrices,
is excluded by the KTeV result. We find $\varepsilon'/\varepsilon
=O(10^{-3})$ for $\cos \theta=1/10$. For instance, we have
$\varepsilon'/\varepsilon =3.3 \times 10^{-3}$ for $m_{3/2}=600$
GeV, $\cos \theta=1/10$ and $\alpha' \simeq \pi/2$. The behaviour
of the $m_{3/2}$-dependence is similar to the both cases of the
first model. \vskip 0.3cm

Furthermore, the CP violation $\varepsilon'/\varepsilon$ can be
reduced if any elements in the Yukawa matrices (\ref{asym-Y})
include suppression factors such that the asymmetry of the
$h$-matrices is reduced. For example, here we replace the (3,1)
element of the down Yukawa matrix as $\lambda^2 \to x\lambda^2$.
For $x=0.1$ we have $\varepsilon'/\varepsilon =O(10^{-3})$ in the
case with $\cos \theta =1/\sqrt{10}$ and $m_{3/2}=O(100)$ GeV.
For instance, we find $\varepsilon'/\varepsilon = 2.5 \times
10^{-3}$ for $x=0.1$, $\cos \theta =1/\sqrt{10}$ and
$m_{3/2}=300$ GeV. \vskip 0.35cm

{\bf 5}. We have studied the CP violation
$\varepsilon'/\varepsilon$ in the models with asymmetric
$A$-matrices as well as asymmetric $h$-matrices. We have shown
that a certain type of the asymmetry enhances
$\varepsilon'/\varepsilon$ and it can be of order of the KTeV
result, $\varepsilon'/\varepsilon \sim 10^{-3}$. A large
magnitude of the asymmetry leads to too large CP violation
$\varepsilon'/\varepsilon$. Thus, we have a constraint on the
large asymmetry of $h$-matrices from the experimental value
$\varepsilon'/\varepsilon \sim 2.8 \times 10^{-3}$.
Other CP violating aspects and FCNC processes would also give further
constraints.
\\ \vskip 0.3cm

The authors would like to thank S.M.Barr for useful discussions.
S.K. would like to acknowledge the support given by the Fulbright
Commission and the hospitality of the Bartol Research Institute.
T.K. is supported in part by the Academy of Finland under the
Project no. 44129.

\end{document}